\documentclass[pra,aps,twocolumn,showpacs]{revtex4-1}

\usepackage{graphicx,amsmath,amsfonts,bm,braket}
\usepackage{hyperref}

\newcommand{\ii}{\mathrm{i}}	        
\newcommand{\ee}{\mathrm{e}}    	
\newcommand{\dd}{\mathrm{d}}		
\renewcommand{\Re}{\mathrm{Re}\,}       
\renewcommand{\Im}{\mathrm{Im}\,}       
\renewcommand{\vec}[1]{\boldsymbol{#1}}	

\begin{document}

\title{Verification of exceptional points in the collapse dynamics
  of Bose-Einstein condensates}

\author{Jonas Brinker}
\author{Jacob Fuchs}
\author{J\"org Main}
\author{G\"unter Wunner}
\author{Holger Cartarius}
\affiliation{Institut f\"ur Theoretische Physik 1, Universit\"at Stuttgart,
  70550 Stuttgart, Germany}
\date{\today}

\begin{abstract}
In Bose-Einstein condensates with an attractive contact interaction
the stable ground state and an unstable excited state emerge in a
tangent bifurcation at a critical value of the scattering length.
At the bifurcation point both the energies and the wave functions of
the two states coalesce, which is the characteristic of an exceptional
point.  In numerical simulations signatures of the exceptional point
can be observed by encircling the bifurcation point in the complex
extended space of the scattering length, however, this method cannot
be applied in an experiment.  Here we show in which way the
exceptional point effects the collapse dynamics of the Bose-Einstein
condensate.  The harmonic inversion analysis of the time signal given
as the spatial extension of the collapsing condensate wave function
can provide clear evidence for the existence of an exceptional point.
This method can be used for an experimental verification of
exceptional points in Bose-Einstein condensates.
\end{abstract}


\maketitle

\section{Introduction}
\label{sec:intro}
In Bose-Einstein condensates with attractive interactions stationary
solutions to the Gross-Pitaevskii equation exist only in certain
regions of the parameter space governing the physics of the condensates.
For the case of an attractive s-wave contact interaction the
condensate collapses when, for given negative scattering length, the
number of particles becomes too large \cite{Dal96,Per97}.
Alternatively, the collapse can be induced experimentally by tuning
the scattering length in the vicinity of Feshbach resonances by
adjusting an external magnetic field \cite{Don01}.
The critical parameter values where collapse occurs correspond to
solutions to the stationary Gross-Pitaevskii equation, where the
stable ground state and an unstable excited state emerge in a tangent
bifurcation \cite{Hue99,Hue03}.
The coalescence of two or even more eigenstates at critical points in
the parameter space, where both the eigenvalues and the eigenvectors
of the states pass through a branch point singularity and become
identical, is a characteristic property of an ``exceptional point''
\cite{Kato66,Moi11}.

Exceptional points cannot occur in quantum systems described by the
linear Schr\"odinger equation with Hermitian operators.
However, they can appear in systems described by non-Hermitian
matrices or in nonlinear systems which depend on a multidimensional
parameter space.
Examples are discussed, e.g., for complex atoms in laser fields
\cite{Lat95}, a double $\delta$ well \cite{Kor03}, the scattering of a
beam of particles by a double barrier potential \cite{Her06},
non-Hermitian Bose-Hubbard models \cite{Gra08}, or models used in
nuclear physics \cite{Bre99}.
The resonant behavior of atom waves in optical lattices \cite{Obe96}
also shows structures originating from exceptional points.
However, the phenomenon of exceptional points in physics is not
restricted to quantum mechanics.
Acoustic modes in absorptive media \cite{Shu00} represent a mechanical
system in which branch-point singularities appear.
Furthermore, manifestations of exceptional points can be seen in
optical devices \cite{Ber94a,Kla08,Wie08}.
The most detailed experimental analysis of exceptional points has been
carried out for the resonances of microwave cavities
\cite{Phi00,Dem03,Die07}, which open the possibility of studying the
properties of the complex resonance frequencies and the wave
functions.

Critical phenomena also occur in nonlinear systems.
Various types of bifurcations which are classified in catastrophe
theory \cite{Pos78} are branch point singularities and resemble
exceptional points in many aspects.
Bose-Einstein condensates are described in a mean-field approach by
the nonlinear Gross-Pitaevskii equation.
The stationary solutions of this equation exhibit a coalescence of two
states due to the nonlinearity of the equation, which turns out to be
a branch-point singularity of the energy eigenvalues and wave
functions \cite{Car08a,Rap09}.
There is only one linearly independent eigenvector of the coalescing
states at the exceptional point.
These systems exhibit the typical consequences of exceptional points,
viz.\ the permutation of eigenstates when an exceptional point is
encircled in the parameter space and a special type of geometric
phase.

For condensates with an attractive gravity-like $1/r$ interaction
\cite{ODe00,Car08a,Car08b} and for dipolar condensates
\cite{Gri05a,Koch08,Lah08a,Lah09,Koe09,Gut13} the bifurcation points
of the ground and excited state have been analyzed in theoretical
computations to verify that they show signatures of exceptional points.
To that aim the stationary states of the nonlinear system can be
approximated by a linear matrix model with a non-Hermitian matrix.
However, the dynamics cannot be described by a linear model because
the superposition principle is not valid in nonlinear systems, i.e.,
there is no unitary time evolution of the condensate wave function.

The stability properties of Bose-Einstein condensates are determined
by the eigenvalues of the Bogoliubov-de Gennes equations which are
obtained by linearization of the Gross-Pitaevskii equation around the
stationary states.
The existence of complex frequencies in the Bogoliubov spectrum
indicates a dynamical instability of the condensate
\cite{Skr00,Kaw04,Kre12,Kre13}.
Typically a stable and an unstable state are created in a tangent
bifurcation, and a state changes its stability properties when running
through an exceptional point at a pitchfork bifurcation.
However, there are counterexamples where the occurrence of an
exceptional point and the stability change do not match.
A discrepancy between the occurrence of a pitchfork bifurcation and
the stability change has been observed in $\mathcal PT$-symmetric
states of a condensate where atoms are incoupled to one side and
extracted from the other \cite{Haa14,Loe14}.

An important property of exceptional points, which follows from the
branch point singularity structure, is the permutation of the
eigenvalues if the exceptional point is encircled in the parameter
space \cite{Kato66}.
Using an analytic continuation of the Gross-Pitaevskii equation the
bifurcation points in Bose-Einstein condensates can be encircled in
the complex plane of the scattering length.
Indeed, after one circle around the critical point a permutation of
the two states is present clearly indicating the existence of an
exceptional point \cite{Car08a,Koe09,Gut13}.
However, complex scattering lengths within the analytically continued
Gross-Pitaevskii equation are not experimentally accessible, and thus
the method mentioned above cannot be used for an experimental
verification of exceptional points in Bose-Einstein condensates.
The aim of this Paper is to present an alternative method for the
verification of the exceptional points.
It is based on dynamical properties of the condensates and can also be
applied in an experiment.

While the investigations of exceptional points mentioned above are
related to non-Hermitian time-in\-de\-pen\-dent Hamiltonians effects of
exceptional points also occur in time-dependent systems
\cite{Lat95,Ste04,Car11,Uzd13}.
Uzdin et al.\ \cite{Uzd13} have shown that a sharp transition from an
oscillatory to a monotonic exponential dynamics occurs in the time
evolution of a single particle in a harmonic trap with a certain
time-dependent frequency $\omega(t)$, and that this transition
corresponds to an exceptional point.
The time evolution of the particle can be analyzed with the harmonic
inversion method as shown in \cite{Fuchs14}.
Here an exceptional point is characterized by the degeneracy of two or
more frequencies in the frequency spectrum of the time signal.

The dynamics of a Bose-Einstein condensate described by the nonlinear
time-dependent Gross-Pitaevskii equation differs fundamentally from
the dynamics of a linear quantum system.
Nonetheless, we will show that the existence of exceptional points can
be verified in the dynamics of the condensate when the harmonic
inversion analysis is applied to a restricted region of the time
evolution.
Starting with a condensate in the stable region of the parameter space
the scattering length can be decreased and then the spatial extension
of the condensate wave function is observed as a function of time.
At the critical scattering length the time signal of the collapsing
condensate exhibits characteristic features in a local time domain
which indicates an exceptional point.

The paper is organized as follows.
The dynamics of Bose-Einstein condensates using a variational approach
to the condensate wave function is discussed in Sec.~\ref{sec:tdvp}.
In Sec.~\ref{sec:analysis} we show that signatures of exceptional
points can be obtained by a local harmonic inversion analysis of time
signals.
The results are presented in Sec.~\ref{sec:results} and conclusions
are drawn in Sec.~\ref{sec:conclusion}.

\section{Variational approach to the condensate dynamics}
\label{sec:tdvp}
In this Paper we investigate Bose-Einstein condensates with an s-wave
contact interaction between particles with mass $m$ in a spherically
symmetric harmonic trap with frequency $\omega$.
With the particle number $N$, the scattering length $a$, and the units
$\ell_{\rm u}=\sqrt{\hbar/(m\omega)}$ for length, $m_{\rm u}=2m$ for
mass, $E_{\rm u}=\hbar\omega/2$ for energy, and $t_{\rm u}=2/\omega$ for
time the dynamics of the condensate wave function is described in a
mean-field approach by the time-dependent Gross-Pitaevskii equation
\cite{Pit03},
\begin{equation}
 \ii\frac{\dd}{\dd t}\psi(\vec r,t) = \left[-\Delta + r^2 + 8\pi Na
   |\psi(\vec r,t)|^2\right] \psi(\vec r,t) \; .
\label{eq:GPE}
\end{equation}
For the wave function we use an ansatz given as superposition of
$N_g$ Gaussian functions,
\begin{equation}
 \psi(\vec r,t) = \sum_{k=1}^{N_g} \exp(-A_k r^2-B_k) \equiv
 \sum_{k=1}^{N_g} g_k \; ,
\label{eq:psi}
\end{equation}
where the $A_k$ and $B_k$ are time-dependent complex variational
parameters.
Variational approaches with coupled Gaussian functions have already
been established in a large variety of applications as a powerful tool
for numerical computations of Bose-Einstein condensates
\cite{Car08b,Rau10a,Rau10b,Rau10c,Eic11,Mar12,Gut13}.
Equations of motion for the variational parameters $A_k$ and $B_k$ are
obtained with the time-dependent variational principle \cite{McL64},
and read
\begin{subequations}
\label{eq:eom}
\begin{align}
 \dot A_k &= -4\ii A_k^2 + \ii V_{2,k} \; ,\\
 \dot B_k &= 6\ii A_k + \ii V_{0,k} \; ,
\end{align}
\end{subequations}
for $k=1,\dotsc,N_g$.
The parameters $V_{0,k}$ and $V_{2,k}$ in Eq.~\eqref{eq:eom} are
solutions of the linear set of equations (for $l=1,\dotsc, N_g$)
\begin{align}
 &\sum_{k=1}^{N_g} \left(\begin{array}{cc}
 \braket{g_l|g_k} & \braket{g_l|r^2|g_k}\\[3pt]
 \braket{g_l|r^2|g_k} & \braket{g_l|r^4|g_k}
 \end{array}\right) \cdot
 \left(\begin{array}{c} V_{0,k}\\[3pt]
     V_{2,k} \end{array}\right)\notag\\
 = &\sum_{k=1}^{N_g}\left(\begin{array}{c}
 \braket{g_l|r^2 + 8\pi Na|\psi|^2|g_k} \\[3pt]
 \braket{g_l|r^4 + 8\pi Nar^2|\psi|^2|g_k}
 \end{array}\right)\,. 
\label{eq:linseteq}
\end{align}
All integrals with the Gaussian functions $g_k$ in
Eq.~\eqref{eq:linseteq} can be solved analytically and are listed in
the Appendix.
For more details of the variational approach and derivations see
\cite{Rau10b}.

\begin{figure}
\includegraphics[width=0.92\columnwidth]{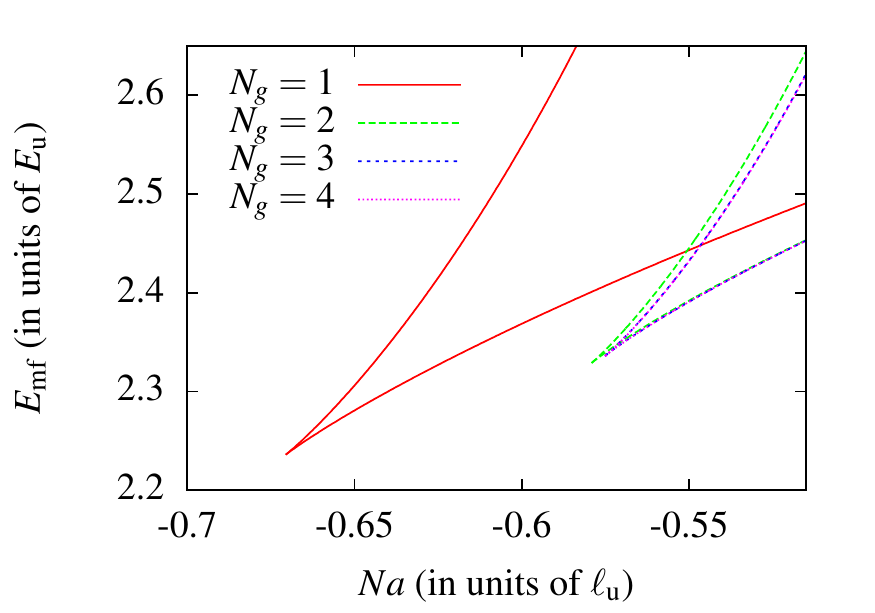}
\caption{(Color online)
  Mean-field energy of the ground and excited state of the BEC as a
  function of the scattering length computed with $N_g=1$ to $4$
  coupled Gaussian functions.  (Results obtained with more than two
  Gaussian functions agree within the line widths.)  The two states
  emerge in a tangent bifurcation at a critical value of the
  scattering length.}
\label{fig1}
\end{figure}

The stationary states of the condensates are obtained by a numerical
root search as fixed points of the equations of motion \eqref{eq:eom}
with the normalization condition $\langle\psi|\psi\rangle=1$.
The mean-field energy of the ground and excited state computed with
$N_g=1$ to $4$ coupled Gaussian functions is presented in Fig.~\ref{fig1}.
As can be seen the mean-field energy converges rapidly with increasing
number of Gaussian functions, and the results obtained with three or
more Gaussian functions fully agree with numerically exact simulations.
The ground and excited state emerge in a tangent bifurcation at a
critical value $(Na)_{\rm cr}$ of the scattering length which depends
on the number of Gaussians $N_g$ used in the computation.
The numerical values of $(Na)_{\rm cr}$ are given in Table~\ref{tab1}.
Like the mean-field energy the critical scattering length converges
rapidly with increasing number of Gaussian functions used in the
computations.
\begin{table}
\caption{Numerical values of the critical scattering lengths of the
  tangent bifurcations computed with one to five coupled Gaussian
  functions.}
\begin{tabular}{c|c}
 $N_g$ & $(Na)_{\rm cr}$~ \\
 \hline
 1 & $-0.67051\,$~ \\
 2 & $-0.57912\,$~ \\
 3 & $-0.57510\,$~ \\
 4 & $-0.574970$ \\
 5 & $-0.574966$
\end{tabular}
\label{tab1}
\end{table}

For the special case of the variational ansatz \eqref{eq:psi} using a
single Gaussian function ($N_g=1$) the linear set of equations
\eqref{eq:linseteq} can be solved in closed form and the equations of
motion \eqref{eq:eom} can be written as (here we drop the index $k$)
\begin{subequations}
\label{eq:eom1}
\begin{align}
 \dot A &= \ii - \frac{8\ii Na}{\sqrt{\pi}}(\Re A)^{5/2} - 4\ii A^2 \; ,\\
 \Im\dot B &= \frac{14Na}{\sqrt{\pi}}(\Re A)^{3/2} +6 \Im A \; .
\end{align}
\end{subequations}
The real part of $B$ is determined by the normalization condition
$\langle\psi|\psi\rangle=1$,
\begin{equation}
 \Re B = \frac{3}{4} \log\left(\frac{\pi}{2\Re A}\right) \; .
\label{eq:ReB}
\end{equation}
In this case the dynamics of the wave function can be obtained as the
canonical equations of the Hamiltonian
\begin{equation}
 H(q,p) = p^2 + \frac{9}{4q^2} + \frac{3\sqrt{3}Na}{2\sqrt{\pi}q^3}
   + q^2 = E_{\rm mf} \; ,
\label{eq:H}
\end{equation}
with the canonical coordinates
\begin{subequations}
\label{eq:qp_coord}
\begin{align}
 q &= \sqrt{\langle r^2\rangle} = \frac{1}{2}\sqrt{\frac{3}{\Re A}} \; ,\\
 p &= -\Im A \sqrt{\frac{3}{\Re A}} \; .
\end{align}
\end{subequations}
It is important to note that the coordinate $q=\sqrt{\langle r^2\rangle}$
describes the extension of the condensate wave function, which means
that the time evolution $q(t)$ can be determined experimentally by
measuring the extension of the wave function.

\section{Local harmonic inversion analysis of time signals}
\label{sec:analysis}
Are the critical scattering lengths in Fig.~\ref{fig1}, where the
ground and excited state of a Bose-Einstein condensate emerge in a
tangent bifurcation, \emph{exceptional points}?
As already mentioned in the introduction this question can be answered
by encircling the critical points in the plane of complex scattering
lengths using an analytic continuation of the Gross-Pitaevskii
equation \eqref{eq:GPE}.
The two states permute after one cycle around the bifurcation point
indicating the existence of an exceptional point, as has been shown for
condensates with long-ranged interactions in \cite{Car08a,Koe09,Gut13}.
Here we want to verify the existence of exceptional points without
resort to the mathematically but not experimentally feasible analytic
continuation of the Gross-Pitaevskii equation.
The idea is to analyze the time evolution of the extension of the
condensate wave function.
The method is first illustrated for the condensate dynamics described
by a single Gaussian function and then extended to the approach with
coupled Gaussians.

\begin{figure}
\includegraphics[width=0.92\columnwidth]{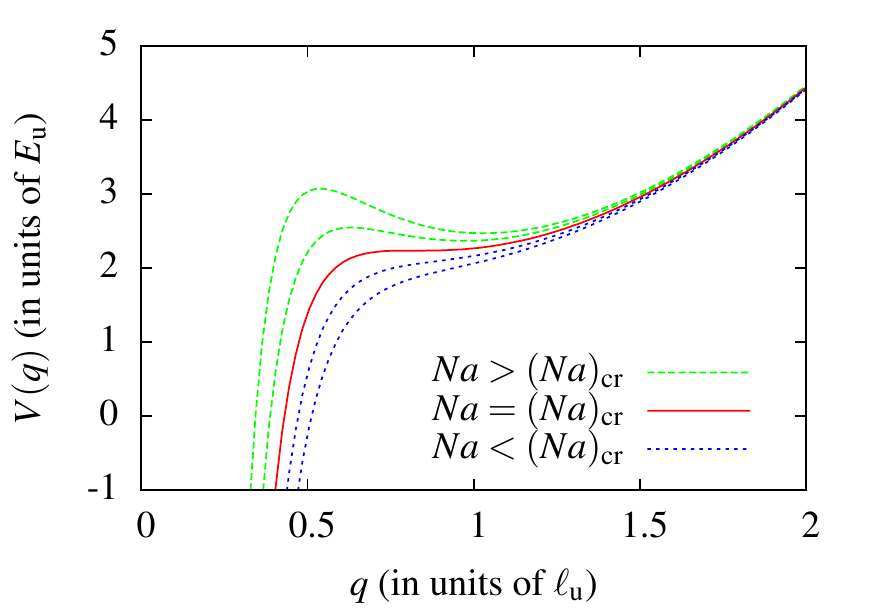}
\caption{(Color online)
  Potential $V(q)$ in Eq.~\eqref{eq:V} for parameters $Na$ above,
  at, and below the critical value $(Na)_{\rm cr}$ of the tangent
  bifurcation.}
\label{fig2}
\end{figure}

As shown in Sec.~\ref{sec:tdvp} the time evolution of the wave
function \eqref{eq:psi} with $N_g=1$ can be described by the
Hamiltonian \eqref{eq:H}, which is effectively the classical dynamics
of a particle in the one-dimensional potential
\begin{equation}
 V(q) = \frac{9}{4q^2} + \frac{3\sqrt{3}Na}{2\sqrt{\pi}q^3} + q^2
\label{eq:V}
\end{equation}
depending on the strength $Na$ of the contact interaction.
The potential $V(q)$ is illustrated in Fig.~\ref{fig2} for scattering
lengths above, at, and below the critical value $(Na)_{\rm cr}=-0.67051$
of the tangent bifurcation.
For $(Na)_{\rm cr}<Na<0$ the potential exhibits a local minimum and
maximum.
These two points characterize a stable and an unstable equilibrium of
the dynamics and thus can be identified as the ground and excited
state of the condensate, respectively.
The two extrema merge at the critical value $Na=(Na)_{\rm cr}$,
thereby forming a saddle with vanishing first and second derivative
in the potential.
For $Na<(Na)_{\rm cr}$ there are no stationary points and the time
evolution $q(t)\to 0$ with increasing time indicates the collapse of
the condensate.

We now show that the exceptional point at the critical value
$Na=(Na)_{\rm cr}$ can be observed in the time evolution $q(t)$ of the
condensate extension.
In what follows we assume that the condensate is initially prepared in
a stationary ground state at scattering length $Na\gtrsim (Na)_{\rm cr}$
and then the scattering length is decreased, in the computation or
experimentally via tuning of a Feshbach resonance, in such a way that
the mean-field energy of the excited state is below the energy of the
initial state or the stationary states do not exist any more.
In that case the condensate collapses, thereby crossing the inflection
point of the potential $V(q)$ with vanishing second derivative.

To simplify the discussion we approximate $V(q)$ in the local vicinity
of the inflection point by the parameter dependent cubic potential
\begin{equation}
 U_\alpha(x) = x^3 - \alpha x \; ,
\label{eq:U}
\end{equation}
where the inflection point has been shifted to the origin.
For $\alpha>0$ two stationary points exist at $x_{1,2}=\pm\sqrt{\alpha/3}$
and degenerate at the critical value $\alpha=\alpha_{\rm cr}=0$.
The nonlinear equation of motion $\ddot x = -3x^2 + \alpha$ for a
particle (with mass $m=1$) moving in the potential \eqref{eq:U}
cannot be solved globally in terms of elementary functions.
For initial conditions $x(0)=0$, $\dot x(0)=v_0$ and short times
the solution can be expanded in a Taylor series
\begin{equation}
 x(t) = v_0 t + \frac{\alpha}{2} t^2 + {\cal O}(t^4) \; .
\label{eq:x_taylor}
\end{equation}
The Taylor expansion \eqref{eq:x_taylor} does not show any special
properties at the critical value $\alpha=0$.
However, the properties of an exceptional point become evident when
$x(t)$ is approximated by a sum of exponential functions,
\begin{equation}
 x(t) \approx \sum_{k=1}^n d_k \exp(-\ii\omega_k t) \; ,
\label{eq:x_hi}
\end{equation}
where the $d_k$ and $\omega_k$ are the amplitudes and frequencies of
the signal, respectively (which both can be complex valued in general).
Using the ansatz \eqref{eq:x_hi} is motivated by the fact that time
signals for systems where the time propagation is described by linear
operators are exactly given by a sum of exponential functions.
The amplitudes and frequencies of a signal \eqref{eq:x_hi} can be
extracted, even for large values of $n$, with the harmonic inversion
method \cite{Wal95,Man97a,Man97b,Mai99d,Bel00,Fuchs14}.
Here we choose $n=2$ which is sufficient to observe the degeneracy of
two frequencies and obtain
\begin{align}
 d_{1,2} &= \pm\frac{\ii v_0^2}{\sqrt{3}\alpha} \quad , \quad
 \omega_{1,2} = \left(\ii\pm\sqrt{3}\right)\frac{\alpha}{2v_0} \; .
\label{eq:x_param}
\end{align}
The Taylor expansion of $x(t)$ in Eq.~\eqref{eq:x_hi} with the
parameters given in Eq.~\eqref{eq:x_param} agrees up to order $t^3$
with the Taylor series in Eq.~\eqref{eq:x_taylor}.
The existence of an exceptional point at the critical value $\alpha=0$
now becomes obvious from the amplitudes and frequencies of the signal
\eqref{eq:x_hi} given in Eq.~\eqref{eq:x_param}.
For $\alpha=0$ the two frequencies coalesce at $\omega_1=\omega_2=0$
and both amplitudes $d_1$ and $d_2$ diverge.
However, in the limit $\alpha\to 0$ the signal \eqref{eq:x_hi}
converges to $x(t)\approx v_0 t$ which can formally be written as
\begin{equation}
 x(t) \approx (\tilde d_0 + \tilde d_1 t) \exp(-\ii\omega t) \; ,
\label{eq:x_hi_degen}
\end{equation}
with the single frequency $\omega=0$ and a prefactor in front of the
exponential function which is a polynomial of degree one in $t$ with
the coefficients $\tilde d_0=0$ and $\tilde d_1=v_0$.
Exceptional points in time signals have been investigated in
\cite{Fuchs14}, where it has been shown that the failure of the ansatz
\eqref{eq:x_hi} due to diverging amplitudes and the occurrence of a
term in the time signal given as the product of a polynomial of degree
$n-1$ in time and an exponential function $\exp(-\ii\omega_k t)$ is a
clear signature of an exceptional point of order $n$.
The frequencies and the coefficients of the polynomials can be
extracted from the signal by the extended harmonic inversion method
developed in \cite{Fuchs14}.
The time evolution of $x(t)$ in Eq.~\eqref{eq:x_hi_degen} thus
indicates the existence of a second order exceptional point.

The analysis of the motion of a particle in the cubic potential
\eqref{eq:U} can now be carried over to the analysis of the
time-dependent extension of a condensate described by the potential
\eqref{eq:V} (see Fig.~\ref{fig2}).
A trajectory $q(t)$ crosses the potential region where the ground
and excited state can coalesce at time $t_0$ determined by the
condition $\dddot q(t_0)=0$.
When that point is shifted to the origin, i.e., $t\to t-t_0$ and
$q(t)\to q(t)-q(t_0)$ the analysis can be performed in the same way as
described above for the cubic potential \eqref{eq:U}.
The difference between the common applications of the harmonic
inversion method to systems with linear time propagators and the
analysis of the nonlinear dynamics of a particle, e.g., in the cubic
potential \eqref{eq:U} is that the time signal in the latter case is
not \emph{globally} given as a superposition of exponential functions,
and therefore the analysis must be restricted to a \emph{local} area
in the time domain.
For the numerical computation of the amplitudes and frequencies we
resort to the harmonic inversion method as introduced in \cite{Fuchs14}.

As explained in Sec.~\ref{sec:tdvp} the description of the condensate
dynamics with canonical coordinates and the Hamiltonian \eqref{eq:H}
is only possible for the simple but not very accurate variational
approach to $\psi(\vec r,t)$ using a single Gaussian function
($N_g=1$) in Eq.~\eqref{eq:psi}.
Using the improved ansatz with coupled Gaussian functions the dynamics
obtained from the equations of motion \eqref{eq:eom} for the
variational parameters $A_k$ and $B_k$ cannot be described by an
effective potential.
However, we can still analyze the time evolution of the extension of
the condensate wave function given as the root of the variance of the
operator $\vec r$, i.e.
\begin{equation}
 q(t) \equiv \sqrt{\langle\psi(t)|r^2|\psi(t)\rangle} \; ,
\label{eq:q}
\end{equation}
which can be expressed in terms of the variational parameters
[see Eq.~\eqref{eq:A2} in the Appendix].
We use the notation $q(t)$ as in Eq.~\eqref{eq:qp_coord} although $q$
is not a canonical variable for coupled Gaussians.
However, it is important to note that the extension of the condensate
as defined in Eq.~\eqref{eq:q} is an observable and can be measured
experimentally.
The time evolution of $q(t)$ obtained with a single Gaussian function
and with coupled Gaussians should qualitatively show a similar
behavior, and thus can be analyzed in the same way using the harmonic
inversion method to verify the existence of exceptional points.
The results are presented and discussed in Sec.~\ref{sec:results}.

\section{Results and discussion}
\label{sec:results}
To analyze the time evolution of the condensate a well defined initial
wave function must be prepared which is not a stationary state of the
Gross-Pitaevskii equation \eqref{eq:GPE}.
A procedure which can be applied theoretically as well as
experimentally is to prepare the stationary ground state of the
condensate for a parameter $Na=s$ of the contact interaction and then
suddenly to change the scattering length $a$.
In an experiment the scattering length can be varied using Feshbach
resonances.
In what follows we present the results of numerical simulations
with a condensate wave function described first by a single Gaussian
function and then with coupled Gaussians.

\subsection{Approach with a single Gaussian function}
\begin{figure}
\includegraphics[width=0.92\columnwidth]{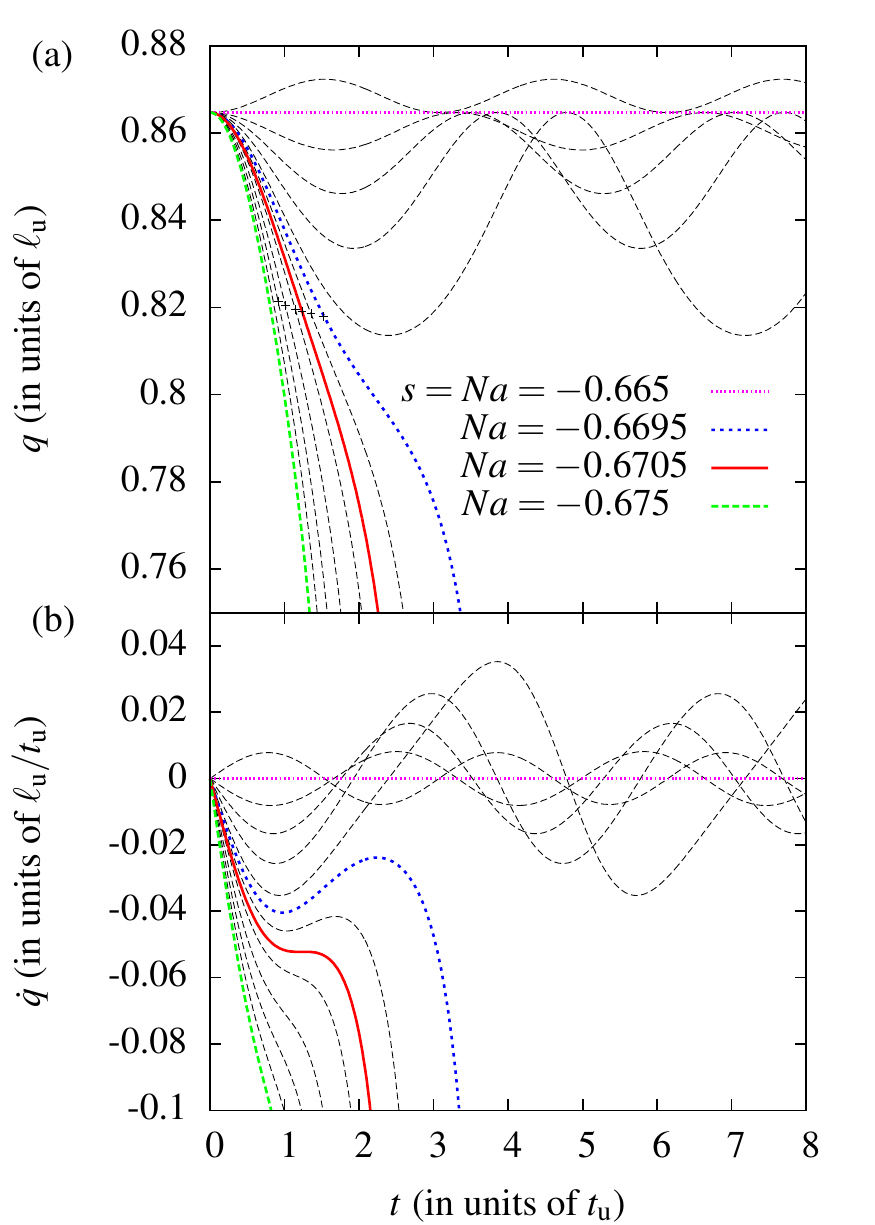}
\caption{(Color online)
  (a) Trajectories $q(t)$ for the time evolution of the condensate
  extension computed with a single Gaussian function, $N_g=1$, and for
  various strengths $Na$ of the contact interaction in the range
  $-0.675\le Na \le -0.664$.  A few trajectories are highlighted and
  labeled.  The initial state of the condensate is prepared at
  $s=Na=-0.665$.  The plus symbols indicate for some trajectories the
  points with vanishing third derivatives $\dddot q(t_0)=0$.
  (b) Time derivatives $\dot q(t)$ of the trajectories in (a).}
\label{fig3}
\end{figure}

In case of an ansatz with a single Gaussian function the ground state
of the condensate for $Na=s$ can be determined as the local minimum of
the potential $V(q)$ in Eq.~\eqref{eq:V}.
After changing the scattering length the equations of motion for the
canonical coordinates $q$ and $p$ are obtained from the Hamiltonian
\eqref{eq:H} and can be integrated numerically.
The trajectories $q(t)$ and the first derivatives $\dot q(t)$ for an
initial state with $s=-0.665$ and various values of $Na$ in the range
$-0.675\le Na \le -0.664$ are presented in Fig.~\ref{fig3}.
If the scattering length is not changed, i.e., $Na=s=-0.665$ then the
extension of the condensate stays constant at $q(t)=0.8647$ in
Fig.~\ref{fig3}(a).
A small change of the scattering length results in a breathing like
dynamics of the wave function, a particle moving in the potential
\eqref{eq:V} (see Fig.~\ref{fig2}) oscillates around the local minimum
but the energy is not high enough to cross the barrier with the local
maximum.
That crossing is possible when $Na$ is reduced below $Na=-0.6693$.
In this case $q(t)$ decreases monotonically and reaches $q=0$ [not
shown in Fig.~\ref{fig3}(a)] within finite time, which indicates the
collapse of the condensate.

Are signatures of the exceptional point visible in Fig.~\ref{fig3}?
When the ground and excited state coalesce in the potential $V(q)$ in
Fig.~\ref{fig2} the dynamics around the critical point is nearly a
free motion linear in time, i.e.,
$q(t)=q(t_0)+v_0(t-t_0)+{\cal O}((t-t_0)^4)$.
The nearly linear behavior can be seen when following the trajectory
with $Na=-0.6705$ in Fig.~\ref{fig3}(a) around $t=t_0=1.22$.
The time derivative $\dot q(t)$ of that trajectory in Fig.~\ref{fig3}(b)
exhibits a saddle indicating the vanishing second and third
derivative of $q(t)$ at $t=t_0$.

\begin{figure}
\includegraphics[width=0.92\columnwidth]{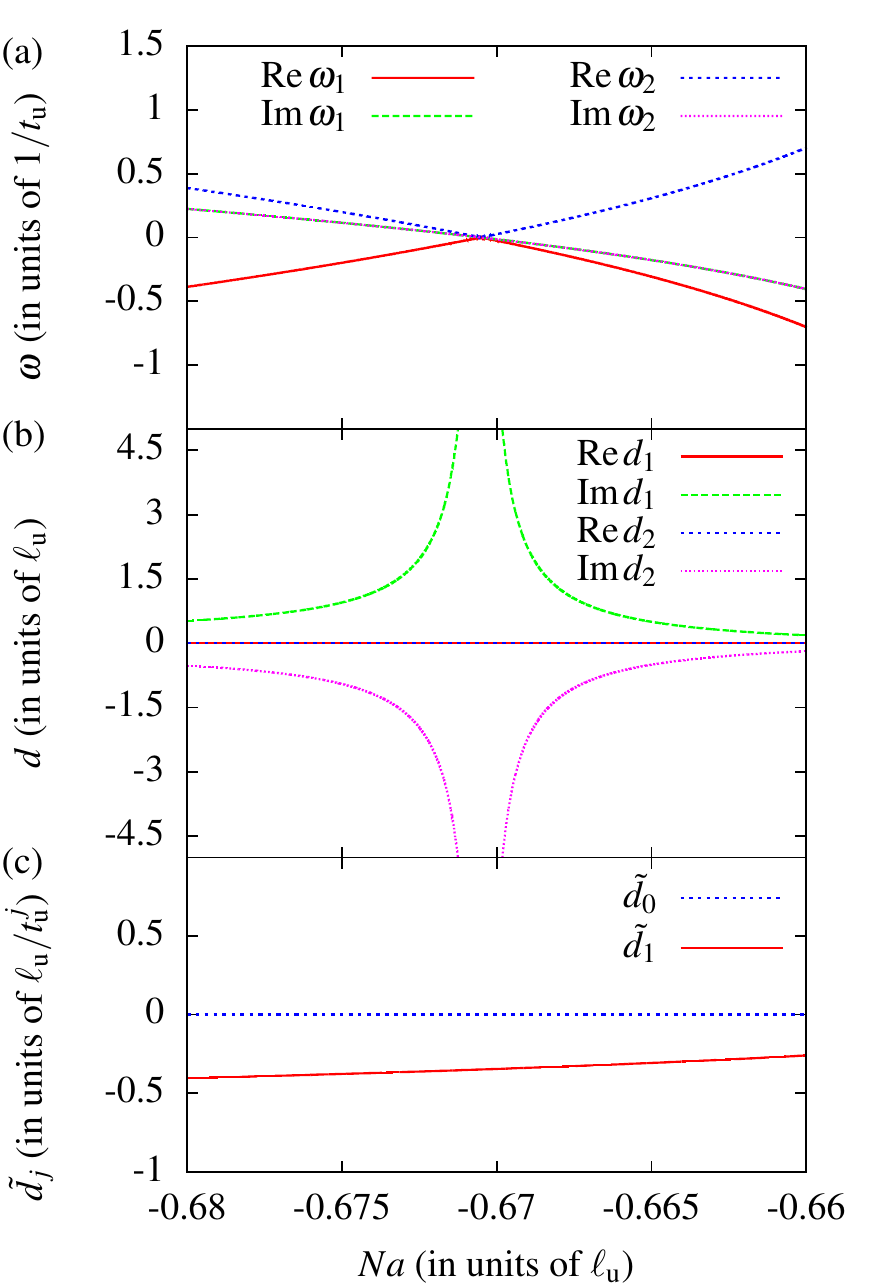}
\caption{(Color online)
  (a) Frequencies and (b), (c) amplitudes obtained by the local
  harmonic inversion analysis of trajectories $q(t)$ computed with a
  single Gaussian function, $N_g=1$, and an initial state prepared
  at $s=Na=-0.57$.  Note that $\Im\omega_1=\Im\omega_2$ and
  $\Re d_1=\Re d_2=0$ and thus the corresponding lines cannot be
  distinguished.  See text for discussion.}
\label{fig4}
\end{figure}

The clear identification of the exceptional point and the precise
determination of the critical scattering length $(Na)_{\rm cr}$ is
possible by analyzing the functions $q(t)$ as described in
Sec.~\ref{sec:analysis}.
For each trajectory the time $t_0$ is computed where $\dddot q(t_0)=0$.
Some of these points are marked by plus symbols in Fig.~\ref{fig3}(a).
For the local harmonic inversion analysis we use only four signal
points $c_j=q(t_0+j\Delta t)-q(t_0)$ with $\Delta t=10^{-3}$ and
$j=0,1,2,3$ to express this signal around $t=t_0$ as the sum of two
exponential functions with two frequencies $\omega_{1,2}$ and
amplitudes $d_{1,2}$.
The results of the local harmonic inversion analysis are presented in
Fig.~\ref{fig4}.
The states have been prepared with the initial parameter $Na=s=-0.57$
and then the time evolution of that state at a modified value $Na$ has
been analyzed.
As can be seen in Fig.~\ref{fig4}(a) the real and imaginary parts of the
frequencies intersect at the critical parameter $(Na)_{\rm cr}=-0.67051$, 
at that point the degenerate frequency is $\omega=0$.
Note that the critical parameter agrees perfectly with the value
given for $N_g=1$ in Table~\ref{tab1}.
The amplitudes $d_1$ and $d_2$ for the harmonic inversion analysis
with an ansatz of two non-degenerate exponential functions are shown
in Fig.~\ref{fig4}(b).
Both amplitudes diverge at $Na=(Na)_{\rm cr}$.
By contrast, the coefficients $\tilde d_0$ and $\tilde d_1$ in a
linear polynomial in $t$ for the ansatz \eqref{eq:x_hi_degen} with a
two-fold degenerate frequency do not show any singularities, as can be
seen in Fig.~\ref{fig4}(c), where the amplitudes $\tilde d_0$ and
$\tilde d_1$ have been computed under the assumption of a single
degenerate frequency $\omega\equiv(\omega_1+\omega_2)/2$.
As outlined in \cite{Fuchs14} the nonzero linear coefficient
$\tilde d_1=v_0$ at the critical parameter $Na=(Na)_{\rm cr}$, where the
two frequencies are degenerate, is the signature of a second order
exceptional point.

\subsection{Approach with coupled Gaussians}
The accuracy of the condensate wave function increases rapidly when
using the ansatz in Eq.~\eqref{eq:psi} with $N_g\ge 2$ coupled
Gaussian functions.
In that case the ground state of the BEC for $Na=s$ is computed
as fixed point of the equations of motion \eqref{eq:eom} with a
numerical root search as described in \cite{Rau10b,Rau10c}.
After changing the scattering length the time dependence of the
variational parameters $A_k$ and $B_k$ is obtained by numerical
integration of the equations of motion \eqref{eq:eom}.
As shown in Fig.~\ref{fig1} and Table~\ref{tab1} the stationary states
and the critical parameter $(Na)_{\textrm{cr}}$ converge rapidly with
increasing number $N_g$ of Gaussian functions.
Our calculations show a similar convergence behavior also for the time
dependence of the trajectories $q(t)$, i.e., for given values $s$ and $Na$
trajectories computed with three or more Gaussian functions are nearly
identical \cite{Bri14}.

\begin{figure}
\includegraphics[width=0.92\columnwidth]{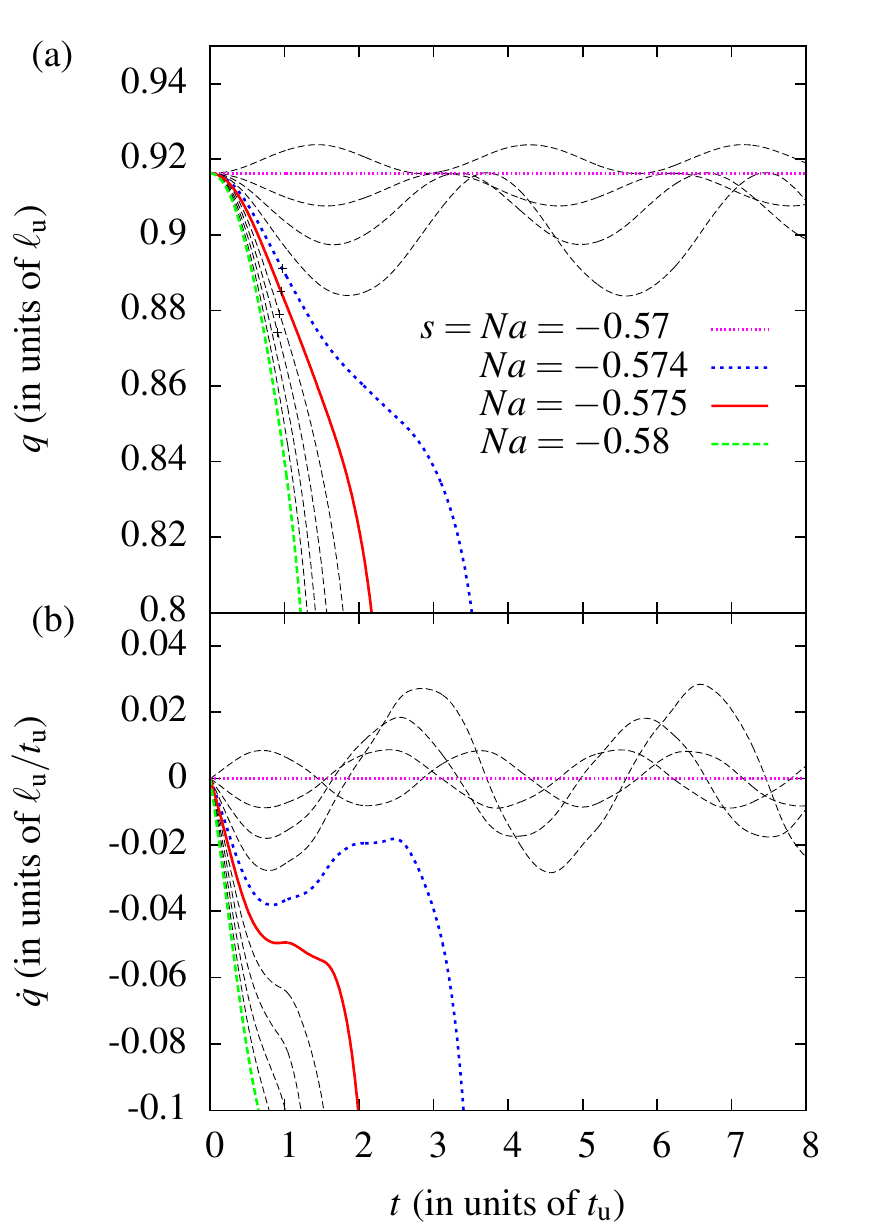}
\caption{(Color online)
  (a) Trajectories $q(t)$ and (b) time derivatives $\dot q(t)$ as in
  Fig.~\ref{fig3} but computed with $N_g=4$ coupled Gaussian functions
  and the initial state prepared at $s=Na=-0.57$.}
\label{fig5}
\end{figure}

For an ansatz with $N_g=4$ coupled Gaussian functions and the
parameter $s=Na=-0.57$ of the initial state the extension of the
condensate as defined in Eq.~\eqref{eq:q} has been computed and the
resulting trajectories $q(t)$ and first derivatives $\dot q(t)$ are
presented in Fig.~\ref{fig5}.
The trajectories $q(t)$ computed with a single Gaussian function in
Fig.~\ref{fig3}(a) and with $N_g=4$ coupled Gaussians in
Fig.~\ref{fig5}(a) qualitatively appear to be very similar, however, 
subtle differences in the derivatives $\dot q(t)$ can be observed when
comparing Figs.~\ref{fig3}(b) and \ref{fig5}(b).
The functions $\dot q(t)$ in Fig.~\ref{fig5}(b) exhibit small
fluctuations with higher frequencies which are absent in
Fig.~\ref{fig3}(b).
The reason is that the dynamics of the condensate when computed with
coupled Gaussian functions does not run exactly along the ``reaction
coordinate'' corresponding to the coordinate $q$ in the
one-dimensional potential \eqref{eq:V} but oscillations in other
degrees of freedom are slightly excited.
Nonetheless, points with vanishing third derivatives $\dddot q(t_0)=0$
can be determined and some of these points are marked by plus symbols
in Fig.~\ref{fig5}(a).

\begin{figure}
\includegraphics[width=0.92\columnwidth]{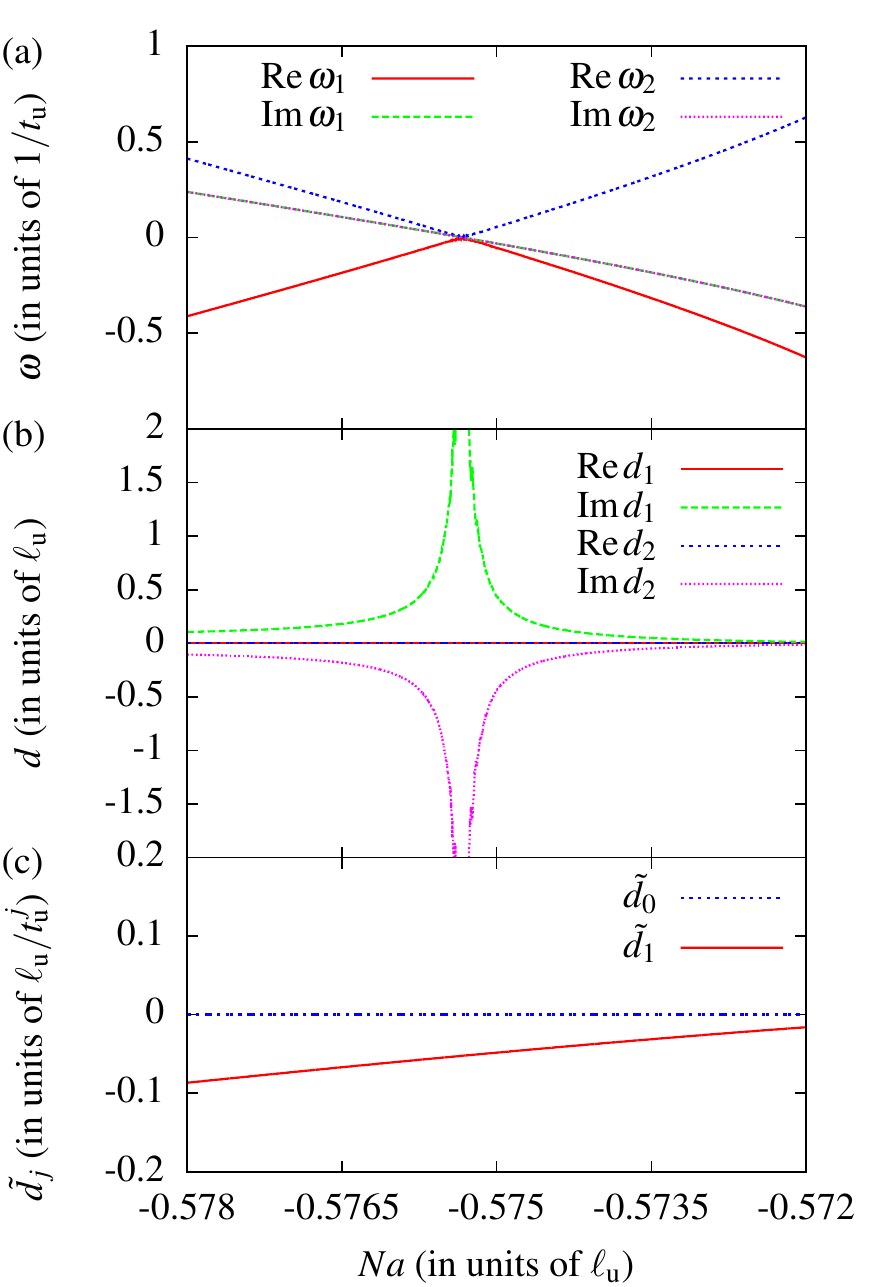}
\caption{(Color online)
  (a) Frequencies and (b), (c) amplitudes obtained by the local
  harmonic inversion analysis of trajectories $q(t)$ as in
  Fig.~\ref{fig4} but computed with $N_g=3$ coupled Gaussian functions
  and the initial state prepared at $s=Na=-0.57$.}
\label{fig6}
\end{figure}

The results of the local harmonic inversion analysis of trajectories
computed with $N_g=3$ coupled Gaussian functions are shown in
Fig.~\ref{fig6}.
The degeneracy of the two frequencies at a critical value
$Na=(Na)_{\rm cr}$ in Fig.~\ref{fig6}(a) and the behavior of the
amplitudes in Fig.~\ref{fig6}(b) and (c) is similar as in
Fig.~\ref{fig4}, however, the exceptional point is shifted to the
critical parameter value $(Na)_{\rm cr}=-0.57534$, which agrees very
well with the value $(Na)_{\rm cr}=-0.57510$ for $N_g=3$ given in
Table~\ref{tab1}.

\begin{figure}
\includegraphics[width=0.92\columnwidth]{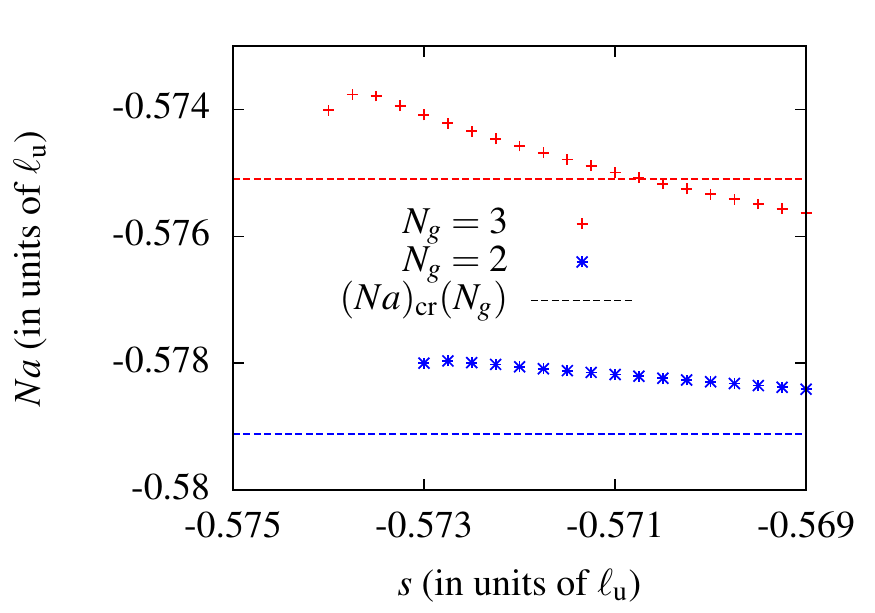}
\caption{(Color online)
  Dependence of the critical parameter $(Na)_{\rm cr}$ on the strength
  $s=Na$ of the contact interaction of the initial state and the
  number of Gaussian functions $N_g$ used for the computations.
  The lines mark the values given in Table~\ref{tab1} for $N_g=2$
  and $N_g=3$.}
\label{fig7}
\end{figure}

We have analyzed the collapse dynamics of condensates initially
prepared at various strengths $s=Na$ of the contact interaction and
computed the critical parameters $(Na)_{\rm cr}$ where the two
frequencies $\omega_{1,2}$ coalesce and the amplitudes $d_{1,2}$ diverge.
For $N_g=1$ the critical value of the exceptional point does not
depend on the initial state, however, for coupled Gaussian functions
the value of $(Na)_{\rm cr}$ slightly depends on $s$.
This can be seen in Fig.~\ref{fig7} for $N_g=2$ and $N_g=3$.
The dependence is caused by the excitation of fluctuations of the
condensate in the various degrees of freedom as discussed above.
However, the analysis of the collapse dynamics with $N_g=3$ coupled
Gaussian functions allows one to determine the position of the
exceptional point as $(Na)_{\rm cr}=-0.575\pm 0.001$, i.e., with an
accuracy of about three significant digits.

The local harmonic inversion analysis of the collapse dynamics can be
applied to verify experimentally the existence of an exceptional point
in a BEC.
In an experiment the initial state of the condensate can be prepared
for a given parameter value $s=Na$ of the contact interaction, and
then the scattering length is quickly ramped to a new value by using
Feshbach resonances.
The variance $\langle \psi(t)|r^2|\psi(t)\rangle$ of the condensate
wave function after a delay time $t$ can be determined with the help
of absorption images.
Since the BEC is destroyed at the snapshot a sufficient number of
identical condensates must be produced and absorption images must be
taken at various delay times $t$ to obtain the time evolution $q(t)$
in Eq.~\eqref{eq:q}.

\section{Conclusion}
\label{sec:conclusion}
The time evolution of Bose-Einstein condensates described in the
mean-field limit by the time-dependent nonlinear Gross-Pitaevskii
equation differs completely from the behavior of quantum systems
obeying the linear Schr\"odinger equation.
Nevertheless, we have shown that signatures of exceptional points
which are well known phenomena in non-Hermitian linear operators can
be observed in the collapse dynamics of Bose-Einstein condensates.
The verification of the exceptional points is possible using the local
harmonic inversion analysis of the time evolution of the condensate
extension during the collapse, and we propose the application of this
method also for an experimental observation of exceptional points in
BECs.

All computations in this Paper are based on the Gross-Pitaevskii
equation obtained with a mean-field approximation.
In calculations with a multi-orbital approach fragmented metastable
states have been observed in the region $Na<(Na)_{\mathrm{cr}}$
\cite{Ced08,Tsa10}.
An interesting question is whether the identification of exceptional
points in this work can be related beyond the mean-field approach to
the splitting of the ground state into multiple fragmented states.

In future studies it will also be interesting to detect critical
phenomena in various time-dependent nonlinear systems by application
of the local harmonic inversion analysis.
The method may even be extended to observe bifurcation points or
higher order exceptional points related to the coalescence of more
than two stationary states in the dynamics of nonlinear systems.

\acknowledgments
This work was supported by Deutsche Forschungsgemeinschaft.\\[1ex]

\appendix

\section{Gaussian integrals}
\label{sec:app}
For the ansatz with coupled Gaussian functions given in
Eq.~\eqref{eq:psi} the integrals required in
Eq.~\eqref{eq:linseteq} read
\begin{align}
\label{eq:A1}
 \braket{ g_k | g_j} &= \frac{\ee^{-(B_j+\bar{B}_k)}\pi^{3/2}}{(A_j+\bar{A}_k)^{3/2}}\,,\\[1ex]
\label{eq:A2}
 \braket{g_k|r^2|g_j} &= \frac{3\ee^{-(B_j+\bar{B}_k)}\pi^{3/2}}{2(A_j+\bar{A}_k)^{5/2}}\,,\displaybreak[1]\\[1ex]
\label{eq:A3}
 \braket{g_k|r^4|g_j} &= \frac{15\ee^{-(B_j+\bar{B}_k)}\pi^{3/2}}{4(A_j+\bar{A}_k)^{7/2}}\,,\displaybreak[1]\\[1ex]
\label{eq:A4}
 \braket{g_k| |\psi|^2|g_j} &= \sum_{l,m=1}^{N_g}
 \frac{\ee^{-(B_j+\bar B_k + B_l + \bar B_m)}\pi^{3/2}}{(A_j+\bar A_k + A_l + \bar A_m)^{3/2}}\,,\displaybreak[1]\\[1ex]
\label{eq:A5}
 \braket{g_k|r^2|\psi|^2|g_j} &= \sum_{l,m=1}^{N_g}
 \frac{3\ee^{-(B_j+\bar B_k + B_l + \bar B_m)}\pi^{3/2}}{2(A_j+\bar A_k + A_l + \bar A_m)^{5/2}}  \,.
\end{align}
The bars indicate the complex conjugate.

%

\end{document}